\documentclass[conference]{IEEEtran}
\IEEEoverridecommandlockouts
\usepackage{cite}
\usepackage{amsmath,amssymb,amsfonts}
\usepackage{algorithmic}
\usepackage{graphicx}
\usepackage{textcomp}
\usepackage{xcolor}
\usepackage{tabularx}
\def\BibTeX{{\rm B\kern-.05em{\sc i\kern-.025em b}\kern-.08em
    T\kern-.1667em\lower.7ex\hbox{E}\kern-.125emX}}
    
\DeclareUnicodeCharacter{2212}{-}
    
\begin{document}

\title{Introduction of Integrated Image Deep Learning Solution and how it brought laboratorial level heart rate and blood oxygen results to everyone\\

\thanks{Identify applicable funding agency here. If none, delete this.}
}

\author{\IEEEauthorblockN{Zhuang Hou}
\IEEEauthorblockA{\textit{Booth School of Business} \\
\textit{University of Chicago}\\
Chicago, United States \\
zhou10@chicagobooth.edu}
\and
\IEEEauthorblockN{Xiaolei Cao}
\IEEEauthorblockA{\textit{Booth School of Business} \\
\textit{University of Chicago}\\
Chicago, United States \\
xcao2@chicagobooth.edu}

}

\maketitle

\begin{abstract}
The general public and medical professionals recognized the importance of accurately measuring and storing blood oxygen levels and heart rate during the COVID-19 pandemic. The demand for accurate contact-less devices was motivated by the need for cross-infection reduction and the shortage of conventional oximeters, partially due to the global supply chain issue. This paper evaluated a contact-less mini-program HealthyPai’s heart rate (HR) and oxygen saturation (SpO2) measurements compared with other wearable devices. In the HR study of 185 samples (81 in the laboratory environment, 104 in the real-life environment), the mean absolute error (MAE) ± standard deviation was 1.4827 ± 1.7452 in the lab, 6.9231 ± 5.6426 in the real-life setting. In the SpO2 study of 24 samples, the mean absolute error (MAE) ± standard deviation of the measurement was 1.0375 ± 0.7745. Our results validated that HealthyPai utilizing the Integrated Image Deep Learning Solution (IIDLS) framework can accurately measure HR and SpO2, providing the test quality at least comparable to other FDA-approved wearable devices in the market and surpassing the consumer-grade and research-grade wearable standards.
\end{abstract}

\begin{IEEEkeywords}
component, formatting, style, styling, insert
\end{IEEEkeywords}

\section{Introduction}
Heart Rate (HR) and oxygen saturation (SpO2) are the most focused measurements during the COVID-19 pandemic. They are both essential metrics providing critical states of a person to determine whether a person is infected by COVID, especially at the early stage of COVID-19 due to the lack of fast test kits. Even if the COVID mortality rate in the US has in general declined following the virus mutation, HR and SpO2 are still consistently tested among the health-focused public. 
Separating people from exposing them to the virus as a strategy to slow down its spread was widely leveraged across the globe. There are numerous strategies to avoid gatherings, such as social distancing, home quarantine, or centralized quarantine. Decentralizing the medical level devices is the key to the strategy’s success. 

The establishing methods on vital signs measurement using smartphone can be split into two categories: Photoplethysmography (PPG)-based \cite{b2, b3} and remote Photoplethysmography (rPPG)-based \cite{b4} methods. With each cardiac cycle, the heart pumps blood to the periphery. With the change of blood volume caused by the pressure, the light absorption is variant. Both methods capture the minor variation of the light reflection and generate a signal for vital signs measurement. The PPG-based method is similar to the principle of some oximeters which requires the users to put their fingers to cover the rear camera with illuminated flashlight. The rear camera can capture the variations of the flow of blood within the vessels. Kanva et al. \cite{b2} measured SpO2 and HR with smartphone rear camera. Nemcovaa et al. \cite{b3} developed an Android application to estimate HR, SpO2 and blood pressure simultaneously.            

The rPPG method applies the same principle of PPG, while the users need to turn on the front camera to capture the face videos. Kwon et al. \cite{b4} built a smartphone application called FaceBEAT that can measure users’ heart rate with face videos. Qiao et al. \cite{b5} built an application that can measure HR and HRV by using the smartphone front camera. Nam et al. \cite{b6} utilized dual cameras to monitor HR and respiration rate. The front camera is applied to capture the motions caused by heartbeats and respirations.

The paper presents a robust workflow, Integrated Image Deep Learning Solution (IIDLS), that realized the laboratory results using real-life quality facial videos. We proved that IIDLS can provide medical level accuracy requiring only on a smartphone. We have tested the framework in the laboratory environment in ensuring the theoretical highest quality results that it can generate and compared the results with the medical chest strap. We have also compared the real-life results with those of the other products, such as the Apple Watch, Fitbit, etc. , claiming offering the same in the markets.   

\section{APPROACH}

Fig. 1. presents the overall flow of the IIDLS estimation framework.
\\
\begin{figure}[htbp]
\centerline{\includegraphics[width=9cm,height=8cm,keepaspectratio]{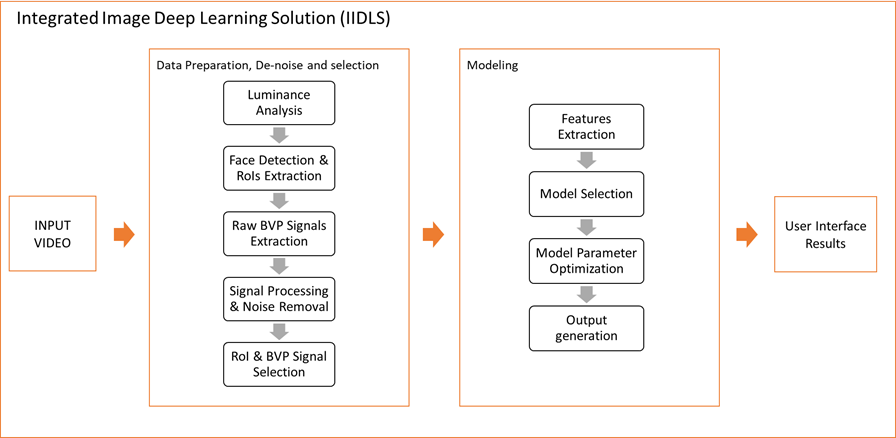}}
\caption{IIDLS estimation framework.}
\label{fig}
\end{figure}
\\
The IIDLS consists of two major steps, data preparation to have the video denoised and the region of interests (RoI) selected, and the main modeling step to create features from the data and predict the outputs. 
The data preparation, the first major step shown in the workflow, extracts candidate BVP signals from sampled multiple face patches. We apply established algorithms to identify the best RoI from the users’ faces that maximize the true signal. 
Once data has been extracted from the RoIs of the facial videos, it will be used as the input to the deep learning algorithms. The model has been trained by 104 videos taken in a real-life environment and 84 videos taken in a laboratory environment preparing for the corresponding comparisons. 

\section{RESULTS}

There are two sets of comparison results depending on the testing environment, in the laboratory environment and the real-life environment. 

\subsection{Laboratory results}

Nine participants, with one female, all of the East Asian descendants and of different ages (20s - 60s) enrolled in the experiment in the laboratory environment. The participants were asked to sit on a chair 1.5m from the camera. The light in the video is consistent from the beginning to the end. There are three 60-second sessions, relax, exercise, and relax sessions, in each 180-second video, where the participants were asked to perform handgrip exercise in the exercise session. The video resolution is VGA (640×480), and the frame rate is 300 fps. Each video is extracted into nine 20-second sequences as the input data fitting the algorithm. Therefore, there are 81 samples for both heart rate and SpO2 comparison. 

Out of all 81 samples, the mean error (ME) between the HR predicted value (through the HealthyPai algorithm) and HR actual value (through the conventional testing devices) is -0.1951 ± 2.2876. The MAE (Mean Average Error) between the HR predicted value (through the HealthyPai algorithm) and HR actual value (through the conventional testing devices) is 1.4827 ± 1.7452.

Compared to polar chest straps, the laboratory results demonstrate that HealthyPai, without any contact, can provide testing results comparable to the polar chest straps that measure heart rate with maximum precision. 

\begin{table}[htbp]
\caption{Heart rate monitor differences from electrocardiogram (mean ± SD)}
\begin{center}
\begin{tabular}{|c|c|c|c|}
\hline
\textbf{Device} & \textbf{\textit{N}}& \textbf{\textit{Paired ME}}& \textbf{\textit{Paired MAE}} \\
\hline
HealthyPai & 81 & -0.2 ± 2.3 & 1.5 ± 1.7 \\
\hline
Polar Chest Strap & 160 & 0.2 ± 1.4 & 0.7 ± 1.2 \\
\hline
\end{tabular}
\label{tab1}
\end{center}
\end{table}

\begin{figure}[htbp]
\centerline{\includegraphics[width=9cm,height=8cm,keepaspectratio]{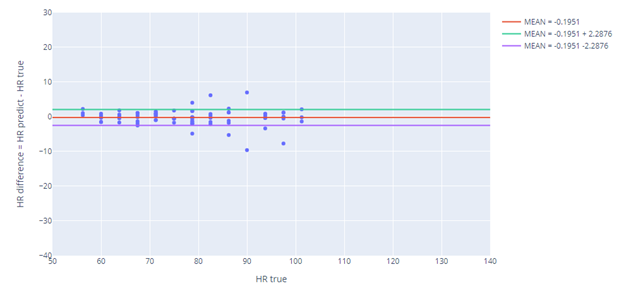}}
\caption{HR difference as HR estimated - HR true value v.s. HR true value}
\label{fig}
\end{figure}

\subsection{Real-life environment results}
In total, there are 25 participants (9 female, 16 male) providing 104 samples in the heart rate comparison and 24 samples (23.08 \% of all samples) in the SpO2 comparison. All 100 \% of the participants are of East Asian descent and taking the videos while resting. 

Out of all 104 samples, the mean error (ME) between the HR predicted value (through the HealthyPai) and HR actual value (through the conventional testing devices) is 0.6538 ± 8.9332. The MAE (Mean Average Error) between the HR predicted value (through the HealthyPai) and HR actual value (through the conventional testing devices) is 6.9231 ± 5.6426. 

HealthyPai provides the quality of results that is at least comparable to other FDA-approved wearable devices and beyond the consumer-grade wearables, which gives an MAE of 7.2 ± 5.4, and research-grade wearables, which gives an MAE of 13.9 ± 7.8. The comparison results are significant through more samples that are collected. 

\begin{table}[htbp]
\caption{Heart rate monitor differences from electrocardiogram (mean ± SD)}
\begin{center}
\begin{tabular}{|c|c|c|c|}
\hline
\textbf{Device} & \textbf{\textit{N}}& \textbf{\textit{Paired ME}}& \textbf{\textit{Paired MAE}} \\
\hline
HealthyPai & 104 & 0.7 $\pm$ 8.9 & 6.9 $\pm$ 5.6 \\
\hline
Apple Watch & 78 & $ −1.7 \pm 10 $ & $5.0 \pm 9.0$ \\
\hline
Fitbit & 80 & 1.0 $\pm$ 8.5 & 5.7 $\pm$ 6.3 \\
\hline
Garmin & 80 & 0.8 $\pm$ 15 & 9.2 $\pm$ 12 \\
\hline
TomTom & 76 & 1.3 $\pm$ 9.8 & 5.4 $\pm$ 8.2 \\
\hline
Consumer-grade wearables &  & & 7.2 $\pm$ 5.4 \\
\hline
Research-grade wearables &  & & 13.9 $\pm$ 7.8 \\
\hline
\end{tabular}
\label{tab1}
\end{center}
\end{table}

\begin{figure}[htbp]
\centerline{\includegraphics[width=9cm,height=8cm,keepaspectratio]{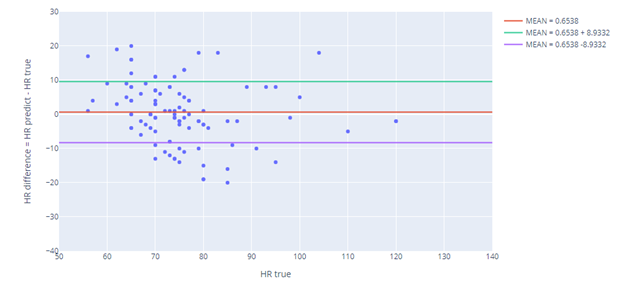}}
\caption{HR difference as HR estimated - HR true value v.s. HR true value}
\label{fig}
\end{figure}

Out of all 24 participants’ results, the mean error (ME) between the SpO2 predicted value (through the HealthyPai) and SpO2 actual value (through the conventional testing devices) is -0.3375 ± 1.2666. The MAE (Mean Average Error) between the SpO2 predicted value (through the HealthyPai) and SpO2 actual value (through the conventional testing devices) is 1.0375 ± 0.7745. 

\begin{figure}[htbp]
\centerline{\includegraphics[width=9cm,height=8cm,keepaspectratio]{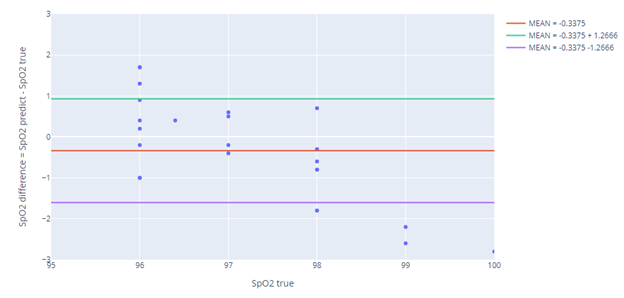}}
\caption{SpO2 difference as SpO2 estimated - SpO2 true value v.s. SpO2 true value}
\label{fig}
\end{figure}

\section{CONCLUSION AND FUTURE WORK}

IIDLS has been thoroughly tested by collecting 185 samples (81 in the lab, 104 in the real-life environment). The comparison shows that IIDLS provides at least comparable quality to other FDA-approved wearable devices in the market and surpasses the consumer-grade and research-grade wearables standards. The environment in which a video was taken is a critical factor affecting the quality of the results. The closer the surrounding environment is to the laboratory, the better quality the user will receive. IIDLS is optimizing the de-noise process to cleanse the signal to the model and has proactively intervened users to either pause the video taking or re-take the videos if the surrounding environment is less optimal.

\vspace{12pt}

\end{document}